\documentclass[aps,prb,twocolumn,superscriptaddress,showpacs,amsmath]{revtex4}

\usepackage{graphicx}
\usepackage{epsfig}

\renewcommand{\vec}[1]{\mathbf{#1}}

\begin{document}
\title{Edge singularities in high-energy spectra of gapped
one-dimensional magnets \\ in strong magnetic fields}
\author{A. Friedrich}
\affiliation{Institut f\"ur Theoretische Physik C, RWTH Aachen, 52056 Aachen, Germany}
\author{A. K. Kolezhuk}
\affiliation{Institut f\"ur Theoretische Physik, Universit\"at Hannover, 30167
  Hannover, Germany}
\affiliation{Physics Department, Harvard University, 17 Oxford Street, Cambridge, MA 02138, USA}
\author{I. P. McCulloch}
\affiliation{Institut f\"ur Theoretische Physik C, RWTH Aachen, 52056 Aachen, Germany}
\author{U. Schollw\"ock}
\affiliation{Institut f\"ur Theoretische Physik C, RWTH Aachen, 52056 Aachen, Germany}

\begin{abstract}
We use the dynamical density matrix renormalization group
technique to show that 
 the
high-energy part of the spectrum of a $S=1$
Heisenberg chain, placed in a strong external magnetic field $H$
exceeding the Haldane gap $\Delta$,
contains edge singularities, similar to those
known to exist in the low-energy spectral response. It
is demonstrated that in the frequency range
$\omega\gtrsim \Delta$
the longitudinal (with respect to the applied field) dynamical
structure factor is dominated by the power-law singularity
$S^{\parallel}(q=\pi,\omega)\propto
(\omega-\omega_{0})^{-\alpha'}$. 
We study the behavior of the high-energy edge exponent $\alpha'$ and the edge
$\omega_{0}$ as functions of the magnetic field.
The existence of edge
singularities at high energies is directly related to
the Tomonaga-Luttinger liquid character of the ground state at
$H>\Delta$ and is expected to be a general feature of one-dimensional
gapped spin systems in high magnetic fields.
\end{abstract}

\date{\today}

\pacs{75.10.Jm, 75.10.Pq, 75.40.Gb, 75.40.Mg}

\maketitle
\section{Introduction}

Studying spectral response  is a
valuable tool to reveal the properties of the strongly correlated ground
state 
in interacting electronic systems. One of the paradigmatic concepts in
physics of one-dimensional (1d) systems is the so-called
Tomonaga-Luttinger liquid (TLL),\cite{Tomonaga-Luttinger} which for 1d systems plays a similar
role as the Fermi liquid theory for higher dimensions.
One of the prominent features of the TLL is the absence of a
quasiparticle peak in the spectral function;  instead, there is a
power-law singularity, with a non-universal 
exponent that depends on the interaction strength.\cite{Schulz86,SachdevSenthilShankar94}

Apart from 1d conductors (e.g., such as carbon
nanotubes\cite{TL-nanotubes1,TL-nanotubes2}) and edge states in
fractional quantum Hall systems,\cite{Wen90,Stone90} the TLL ground
state is expected to exist in several other 1d systems, particularly
in spin chains and ladders. For an antiferromagnetic (AF)
$S=\frac{1}{2}$ Heisenberg chain, the TLL character of the ground
state can be rigorously established from the Bethe ansatz
solution,\cite{Haldane80-81} and it is believed
\cite{Haldane83,AffleckHaldane87} that this picture is valid also for
the other gapless AF chains with half-odd-integer $S$.

For AF Heisenberg chains with integer $S$ and for spin ladders, which
have a finite excitation gap $\Delta$, it has been argued by
analytical \cite{ChitraGiamarchi97,KonikFendley02} and numerical
\cite{Campos02,Fath2003} methods that in an external magnetic field
$H>\Delta$, strong enough to close the spin gap and cause a finite
magnetization, the ground state is also of the TLL type.

Recently, in a search for Luttinger liquid signatures in spin systems,
several experimental studies have been undertaken.
\cite{Regnault+06,Hagiwara+05,Hagiwara+05a,Izumi+03,Yoshida+05} Most
of the experimental evidence is, however, indirect, based, e.g., on
the analysis of the temperature dependence of specific heat
\cite{Hagiwara+05a,Yoshida+05} or NMR relaxation rate.\cite{Izumi+03} 
Direct detection of the low-energy excitation
continuum in inelastic neutron scattering experiments
\cite{Regnault+06,Hagiwara+05} is a difficult task, not to mention
extracting the dynamical exponents from the low-energy spectrum.
High-energy modes might be easier to study, especially with techniques such 
as electron spin resonance (ESR).\cite{Orendac+99,Hagiwara+03}

At the same time, the high-energy excitations can be viewed as mobile impurities
interacting strongly with the underlying TL liquid, and their spectrum can bear
similar features as those found in TLL, namely the absence of the quasiparticle
peak which is replaced by an edge
singularity.\cite{SorellaParola96,FurusakiZhang99,KM02prb,KM02ptp} It was shown
\cite{KM02prb} that the spectrum of high-energy excitations with $S^{z}=0$, $1$
contains an edge singularity with a nontrivial field dependence of the edge
frequency which in the idealized model with no interaction except the hardcore
constraint is given by $\omega_{0}=(1-S^{z})H$. The spectral function of a
mobile impurity in the TL liquid has been extensively studied theoretically;
\cite{NetoFisher96,Tsukamoto98} however, to our knowledge, no numerical results
are available to compare with the theoretical predictions. An alternative
bosonization description including high-energy modes has been proposed
recently.\cite{Sato06}  All of this
motivates further study of the high-energy spectra of 1d gapped spin systems in
strong field. 

In the present paper, we study the $S=1$ Heisenberg chain, which is a paradigmatic
example of a 1d gapped antiferromagnet. There is a large body of numerical and
theoretical results concerning its low-energy behavior in strong fields
\cite{KonikFendley02,Campos02,Fath2003,Affleck05} which can be used for
consistency checks.  Using the dynamical density matrix renormalization group
(DMRG) technique, we will show that the high-energy part of the spectrum of a
$S=1$  chain, placed in a strong external magnetic field $H$ exceeding
the Haldane gap $\Delta$, contains edge singularities, similar to the well-known
infrared singularities in the low-energy spectral response. It will be
demonstrated that in the frequency range $\omega\gtrsim \Delta$ the longitudinal
dynamical structure factor is dominated by the continuum with a power-law edge
singularity $S^{\parallel}(q=\pi,\omega)\propto (\omega-\omega_{0})^{-\alpha'}$.
The edge exponent $\alpha'$ is found to decrease as a function of magnetization,
and the edge $\omega_{0}$ is shown to follow approximately the linear law
$\omega_{0}\simeq H$ as found earlier.\cite{KM02prb,KM02ptp,FurusakiZhang99} 

\section{Theoretical preliminaries}

Consider the $S=1$ Heisenberg chain in an applied field described by the
Hamiltonian
\begin{equation} 
\label{s1ham}
\widehat{\mathcal{H}}=J\sum_{n}\vec{S}_{n}\cdot\vec{S}_{n+1} -H\sum_{n}S^{z}_{n},
\end{equation}
where $\vec{S}_{n}$ is the spin-$1$ operator at site $n$,  the
exchange constant $J$ will be set to unity, and the external magnetic
field $H$ is applied along the $z$ axis. In absence of the applied
field, the ground state of the model is a singlet, and the lowest
excitations are the triplet of Haldane magnons separated from the
ground state by the gap $\Delta\simeq 0.41$ at the wave vector $q=\pi$. 
The gap 
closes for $H>H_{c}=\Delta$, and the system acquires a finite density of
$S^{z}=+1$ magnons, which can be viewed
as bosonic particles satisfying the hardcore constraint. It is
convenient to redefine the momentum so that the minimum of the magnon
dispersion will correspond to zero.
Interacting $S^{z}=1$ magnons form a TL liquid with
the Hamiltonian given by
\begin{equation}
\label{TLham}
\mathcal{H}_{0} =  \frac{v_{F}}{2}\int dx \, \Big\{\frac{1}{K}(\partial_x \varphi)^{2} 
+ K (\partial_x \theta)^{2}\Big\},
\end{equation}
where $\varphi$ and $\theta$ is a pair of bosonic fields satisfying the
commutation relations $[\varphi(x),\theta(x')] = i\Theta (x'-x)$ (here
$\Theta(x)$ is the Heaviside function) and connected by the duality relation
$\partial_{t}\varphi = v_{F} \partial_{x}\theta$. The physical properties of the
TL liquid are
completely characterized by its Fermi velocity $v_{F}$ and
the TLL parameter $K$. Both $v_{F}$ and $K$ are functions of $H$, and
their behavior for the Haldane chain has been studied analytically
\cite{KonikFendley02,Affleck05} as
well as numerically.\cite{Campos02,Fath2003} An important feature of the
effective TLL description of the Haldane chain is the fact that the
parameter $K$ is always larger than unity ($K\to1$ as $H\to H_{c}$).

The local spin density $\rho(x)$
(i.e., the density of the sea particles) can be expressed as
\begin{equation} 
\label{Sz} 
\rho=S^z=m+\frac{1}{\sqrt{\pi}}\partial_x \varphi
+\mbox{const}\times \sin \big\{2k_Fx+ \sqrt{4\pi } \varphi\big\} ,
\end{equation}
where $m=m(H)\equiv k_{F}/\pi$ is the equilibrium magnetization density at the
given field.  The operator $a^{\dag}\sim e^{i\sqrt{\pi} \theta}$ creates a kink
in the $\varphi$ field shifting it by $\sqrt{\pi}$ and thus corresponds to the
creation of a $S^{z}=1$ particle, so that at low energies one can establish the
correspondence $S^{+}\sim a^{\dag}$. The low-energy contribution to the
transversal dynamical structure factor (DSF) is thus given by
$S^{+-}(q=\pi,\omega)\propto \int dx\int dt \, e^{i\omega t}\langle
a^{\dag}(x,t)a(0,0)\rangle$, and, since
\[
\langle a^{\dag}(x,t)a(0,0)\rangle\propto
(v_{F}^{2}t^{2}-x^{2})^{-\eta/2}\quad \mbox{with}\quad
\eta=\frac{1}{2K},
\]
one readily obtains the infrared singularity 
\begin{equation} 
\label{dsf-low}
S^{+-}(q=\pi,\omega)\propto \frac{1}{\omega^{\alpha}},\quad \alpha=2-\eta 
\end{equation}
with the edge exponent $\alpha$ determined solely by the TL liquid
parameter $K$.

To describe the excitation of ``impurity'' particles corresponding to
the higher-energy magnon
branch with $S^{z}=0$, one can introduce another bosonic field $b(x)$
described by the Hamiltonian
\begin{equation} 
\label{Hb}
\mathcal{H}_{b}=\int dx \Big\{ \omega_{0} b^{\dag}b 
+\frac{1}{2M}(\partial_{x}b^{\dag})(\partial_{x}b)\Big\}.
\end{equation}
The scaling dimension of the field $b$ is ${\rm
dim}[b]=\frac{1}{2}$. The TLL action is Lorentz-invariant and thus
dictates the dynamical exponent $z=1$, from which one can see that the
scaling dimension of the mass $M$ is equal to $+1$, i.e., $M$ flows to infinity
so that one can neglect the
``impurity'' dispersion described by the second term in (\ref{Hb});
this argument is essentially identical to the dynamical localization of a hole
moving in an antiferromagnet as discussed in Ref.\
\onlinecite{SachdevTroyerVojta01}.

We will be interested in the contribution of the excited ``impurity''
particles ($b$-particles) to the dynamical structure factor
$S^{zz}(q=\pi,\omega)$, which will be determined by
$b$-particles with zero momentum. The local interaction between 
the ``impurity'' particles and the sea particles
($a$-particles) is proportional to the product of corresponding densities 
$b^{\dag}b \rho$.\cite{remark-on-current} This yields, apart from
renormalizing the energy $\omega_{0}\mapsto
\omega_{0}+\mbox{const}\cdot m$, two interaction terms: one
corresponds to
forward scattering and is given by
\begin{equation} 
\label{V-forw} 
\mathcal{H}_{int}^{f}=U_{f}\int dx \frac{1}{\sqrt{\pi}}
(\partial_{x}\varphi) b^{\dag}b,
\end{equation}
and the other, which is proportional to $b^{\dag}b\sin\big\{2k_Fx+
\sqrt{4\pi } \varphi\big\}$, describes backscattering. The
backscattering term, however, is irrelevant  since its scaling
dimension is $1+K$, and for the Haldane
chain one always has $K>1$. With the remaining interaction
(\ref{V-forw}), the Hamiltonian
can be diagonalized by means of a unitary transformation
\cite{AffleckLudwig94}
\begin{equation} 
\label{Ut} 
\widehat{U}=e^{\displaystyle -i\delta \int dx \theta(x) b^{\dag}(x)b(x)},\quad
\text{with} \quad \delta=\frac{K U_{f}}{v_{F}\sqrt{\pi}},
\end{equation}
which changes the phase of $b$ operator according to
\begin{equation} 
\label{UbU} 
b\mapsto \overline{b}=\widehat{U}^{\dag} \,b\, \widehat{U}=b\, e^{-i\delta \theta }.
\end{equation}
Apart from removing the interaction (\ref{V-forw}), the unitary
transformation (\ref{Ut}) just renormalizes the threshold energy
$\omega_{0}$, which thus can in principle depend both on the interaction and
on the applied field. 

Exciting a $b$-particle requires ``disturbing'' the sea of $a$-particles,
because the hardcore constraint leads to a change in the allowed momentum values
when the total number of particles is changed.\cite{KM02prb} To take the
hardcore nature of the bosons into account, we can postulate that the physical
creation operator of the $b$-boson can be represented as
\begin{equation} 
\label{psi-b}
\psi_{b}^{\dag}=b^{\dag}a^{\dag}=\overline{b}^{\dag}e^{i(\sqrt{\pi}-\delta)\theta}.
 \end{equation}
With that definition, $b$ is the ``color-changing'' operator for an
existing sea particle, and in absence of any other interaction except
the hardcore constraint the excitation of a $b$-particle would be
equivalent to adding another sea boson and the only difference would
be the additional energy cost $\omega_{0}$.

We are interested in the correlator 
$\langle
\psi_{b}^{\dag}(x,t) \psi_{b}(0,0)\rangle$
because it determines the ``impurity'' contribution to the longitudinal DSF
\begin{equation} 
\label{dsf1} 
S^{zz}(\pi,\omega)\propto \int dx \int dt e^{i\omega t} \langle
\psi_{b}^{\dag}(x,t) \psi_{b}(0,0)\rangle
\end{equation}
Since the $\overline{b}$ quasiparticles are free after the
transformation, the correlator factorizes. One has
$\langle \overline{b}^{\dag}(x,t) \overline{b}(0,0)\rangle\propto
\delta(x)\,e^{-i\omega_{0}t}$, where the Dirac delta-function $\delta(x)$ is the
consequence of irrelevance of the dispersion term in (\ref{Hb}), so that
\begin{equation} 
\langle \psi_{b}^{\dag}(x,t) \psi_{b}(0,0)\rangle \propto 
e^{-i\omega_{0}t} \delta(x) /t^{\eta'},
\end{equation}
where
\begin{equation} 
\label{eta1} 
\eta'=\frac{1}{2K}\Big( 1-\frac{KU_{f}}{\pi v_{F}}\Big)^{2}
\end{equation}
is the ``impurity exponent''. The longitudinal dynamical structure factor thus
contains an edge singularity 
\begin{equation} 
\label{dsf2} 
S^{zz}(\pi,\omega)\propto
\frac{1}{(\omega-\omega_{0})^{\alpha'}}\quad\mbox{with}\quad \alpha'=1-\eta'.
\end{equation}
The edge exponent $\alpha'$, in contrast to the low-energy edge
exponent $\alpha$, depends not only on the TL liquid parameter $K$,
but also on the Fermi velocity $v_{F}$ and on the ``host-impurity'' interaction
$U_{f}$. In the noninteracting hardcore case, which corresponds to $U_{f}=0$ and
$K=1$, this exponent takes the
value $\alpha'=\frac{1}{2}$, in agreement with earlier
studies.\cite{KM02ptp,KM02prb,FurusakiZhang99} 
 
Comparing (\ref{dsf2}) and (\ref{dsf-low}), one can see that the high-energy DSF
exponent $\alpha'=1-\eta'$ should be considerably smaller than the low-energy
one $\alpha=2-\eta$. In our approach, this is a direct consequence of the
dynamical localization of the ``impurity'', which in turn followed from the
non-relativistic form of the $\mathcal{H}_{b}$ Hamiltonian (\ref{Hb}).
It is worthwhile to note that the same answer is obtained assuming that
$\mathcal{H}_{b}$ has a Lorentz-invariant form and describes massive field with
mass $\omega_{0}$ and characteristic limiting velocity  $c$. In this latter case $\langle
\overline{b}^{\dag}(x,t) \overline{b}(0,0)\rangle\propto K_{0}(\omega_{0}r/c)$,
where $K_{0}$ is the Bessel function, and $r=\sqrt{x^{2}-c^{2}t^{2}}$, and one
can show\cite{FurusakiZhang99} that the corresponding Fourier transform yields
the same asymptotic behavior (\ref{dsf2}).

\section{Technique}

We use the density matrix renormalization group (DMRG) method
\cite{White1992-11, Schollwoeck2005} to calculate the spectral
function defined earlier. The DMRG works by truncating the Hilbert space
of the system, based on selecting the optimal states from the
Schmidt decomposition of the lattice as split into two (left and right) blocks.
The method we use is a matrix product generalization of the 
``correction vector method'' \cite{Kuehner1999}, in which we calculate the
Green's function
\begin{equation}
G(q,\omega+i\delta)= \langle 0|\widehat{A}^{\dagger}_q
\frac{1}{E_0+\omega+i\delta-\widehat{H}}\widehat{A}^{\phantom{\dagger}}_q| 0\rangle
\end{equation}
at a given frequency $\omega$ with $E_0$ being the ground state energy
of the system and $\delta$ a numerical broadening,
via the so-called ``correction vector'',
\begin{equation}
|c(w+i\delta)\rangle = (E_0 + w + i \delta - \widehat{H})^{-1}\widehat{A}_q |0\rangle,
\end{equation}
which is determined by a standard linear solver (we used GMRES, but biconjugate
gradient is also popular).  In contrast to the traditional dynamical DMRG
approach where the groundstate $|0\rangle$, Lanczos vector $\widehat{A}_q
|0\rangle$ and correction vector are all determined simultaneously, the matrix
product formulation allows the calculation to be split up, improving both the
speed and accuracy (for a given size of the truncated Hilbert space) of the
calculation.  We first calculate the matrix product approximation to the
groundstate $|0\rangle$ by a standard DMRG calculation, so that we can apply
$\widehat{A}_q$ exactly to give the Lanczos vector, which is independent of
frequency.  This is used as input to the correction vector solver which is
trivially parallelizable over different frequencies, but we can also e.g., use a
previously determined correction vector from a nearby frequency as the initial
''guess'' vector, additionally accelerating the calculation. The matrix product
formulation provides good control over the errors, as the residual norm $|| \;
\widehat{A}_q |0\rangle - (E_0 + w + i \delta - \widehat{H})|c(w+i\delta)\rangle
\; ||$ can be calculated \textit{exactly}. This is prohibitively difficult in
the standard DMRG approach as it requires calculating the matrix elements of
$\widehat{H}^2$ and the usual dynamical DMRG approximation $\widehat{H}^2 \simeq
(P\widehat{H}P^\dagger)^2$, where $P$ is the projector onto the truncated
Hilbert space, is inadequate and produces an estimate for the residual norm that
is many orders of magnitude too small.  Details of this calculation will be
provided elsewhere (see also reference [\onlinecite{VerstraeteMPS}] for a
similar approach to this calculation).  Given the correction vector, the Green's
function can be obtained via
\begin{equation}
G(q,\omega+i\delta)= \langle \widehat{A}_q^\dagger |c(\omega+i\delta)\rangle,
\end{equation}
from which the spectral function
\begin{equation}
S(q, \omega) = - \frac{1}{\pi} \lim_{\delta \rightarrow 0^+} \text{Im}\, G(q, \omega)
\end{equation}
can be obtained.

\section{Results and discussion}

\begin{figure}
\epsfig{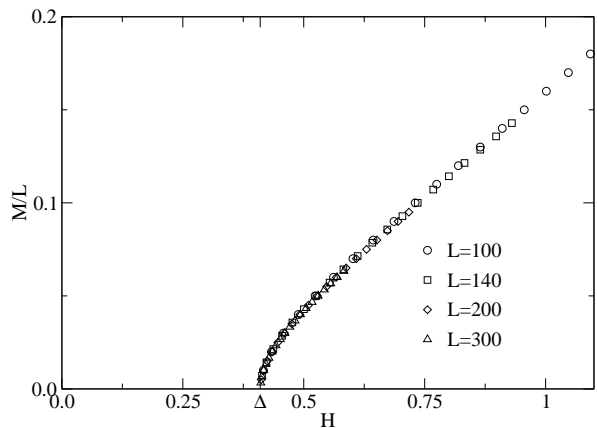}
\caption{\label{fig:magn_H}
Ground state magnetization per site versus magnetic field for different system sizes.}
\end{figure}

We start with the low-energy properties of the transversal 
DSF, which are better understood analytically, to show the validity of
the method. 
First of all, we have to determine the total spin of the ground state
depending on the applied field strength.
In Fig.~\ref{fig:magn_H} we show the ground state magnetization $M$ as a function of 
the magnetic field $H$; the results for different system sizes are
consistent, indicating that finite size effects in $M(H)$ are small for $L\gtrsim 100$. 

\begin{figure}
\begin{center}
\epsfig{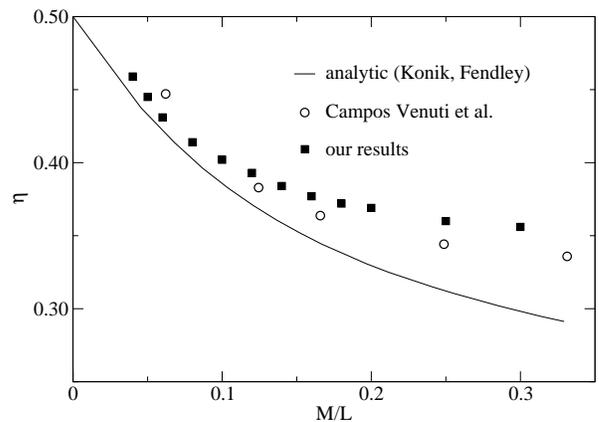}
\end{center}
\caption{Exponent $\eta$ of the static correlation function. Solid line: results
obtained from analytic calculations by Konik and Fendley\cite{KonikFendley02}
 based on the effective
nonlinear sigma-model description.
Open circles: DMRG results for short chain lengths $L<80$ obtained by Campos
Venuti {\it et al.} \cite{Campos02} Filled squares: our results obtained from
DMRG calculations using chains of length $L=200$.}\label{fig:eta}
\end{figure}

The exponent $\eta$ can be obtained from the static transverse spin-spin
correlation function \mbox{$\langle S^x(x)S^x(0) \rangle \sim |x|^{-\eta}$}
which is easily accessible in DMRG calculations. From this the TLL parameter can
be calculated via $K=1/(2\eta)$. The values for $\eta$ are shown In
Fig.~\ref{fig:eta} we show our values for $\eta$  and compare 
them with analytical results based on the effective
description in the nonlinear sigma-model framework\cite{KonikFendley02} as well
as with the earlier DMRG results.\cite{Campos02} From this $\eta$ we get a first estimate
for the edge exponent $\alpha$ via the second equation in Eq.\ (\ref{dsf-low}).

\begin{figure}
\begin{center}
\epsfig{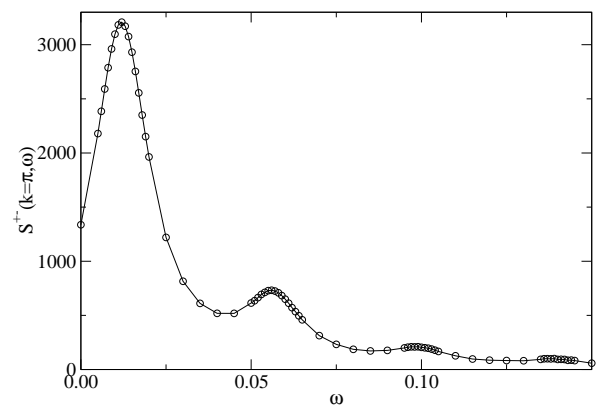}
\end{center}
\caption{Spectral function $S^{+-}(k=\pi, \omega)$ for a $200$-site chain at
magnetic field $H=0.54$ leading to a ground state magnetization per site of
$M/L=0.05$. Here, $\delta = 0.01$ and $m=300$ states were kept in the
calculation.}\label{fig:lower_cont}
\end{figure}

A typical scan of the low energy-continuum of $S^{+-}(k=\pi, \omega)$ is shown in
Fig.~\ref{fig:lower_cont} for a $200$-site chain. The magnetic field applied
is $H=0.54$ in units of the coupling constant leading to a ground
state magnetization per site of $M/L=0.05$.

\begin{figure}
\begin{center}
\epsfig{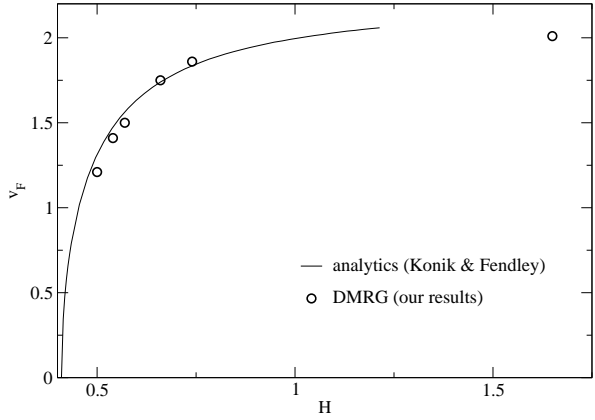}
\end{center}
\caption{Fermi velocity extracted from $S^{+-}(k=\pi,\omega)$ as a function of
magnetic field (open circles). The solid line corresponds to Eq.(40) of Konik
and Fendley \cite{KonikFendley02} with the bare spin wave velocity 
 $v_0 = 2.5$.}\label{fig:vF}
\end{figure}

The spectrum reveals a set of peaks at discrete frequency values, as one expects
for a finite-size system. The peaks are nearly equidistant (in fact, the
distance between the peaks decreases slightly with the frequency, which is
naturally explained by the effective decrease of the Fermi velocity as the Fermi
sea gets emptied by more and more particles being excited). From the energy
difference of the first two peaks we extract the Fermi velocity $v_F$; its
dependence on the magnetic field is in
good agreement with the results  of Konik and Fendley \cite{KonikFendley02} (see
Fig.~\ref{fig:vF}) if one sets their parameter $v_{0}$ to $2.5$. This parameter
has the meaning of a
bare spin velocity in the nonlinear sigma model description of Konik and
Fendley, and the value $v_{0}\approx 2.5$ agrees well with the known spin wave
velocity $v\approx 2.46$ of $S=1$ Haldane chain.\cite{White92}

One might be tempted to extract the edge exponent $\alpha$ from the dependence
of the peak heights in the spectral function (Fig.~\ref{fig:lower_cont}) on the
frequency $\omega$. However, this is not a very good idea: although all these
peaks are numerically broadened with the same constant $\delta=0.01$ for a
$L=200$ site chain, they do not all show the same width as it would have been
expected. With inceasing frequency they get broader. This can be understood in
the following way: The higher the excitation energy is, the more combinations
exist to reach it by exciting several particle-hole pairs. In a system with an
ideal linear dispersion the energies of all those combinations would be exactly
the same and equal $2\pi v_{F} n/L$ with $n$ being some integer. In a real
system the dispersion is not exactly linear, and those energies become slightly
scattered around the $2\pi v_{F} n/L$, contributing an additional broadening to
peaks with higher energy. Hence the height of the higher energy peaks is
underestimated and thus the decay is overestimated leading to a value for the
exponent $\alpha$ being too large.

\begin{figure}
\begin{center}
\epsfig{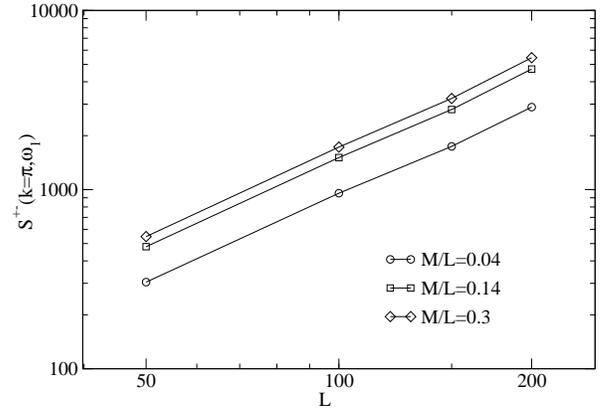}
\end{center}
\caption{Peak height of the first peak in $S^{+-}(k=\pi, \omega)$ as function of
the system size $L$ for different magnetizations. Fitting this log-log plot by a
straight line yields the edge exponent $\alpha$. }\label{fig:peakheight_L}
\end{figure}

There is a much better way to determine the edge exponent, namely from
the size dependence of the peak strength.
The height of the first peak at $\omega_1$ is given by
\[
S^{+-}(k=\pi, \omega_1) \propto 1/\omega_1^\alpha.
\]
With $\omega_1\propto v_{F}/L$ we find
\[
  S^{+-}(k=\pi, \omega_1) \propto L^\alpha
\]
for fixed magnetization per site.  This is shown in a log-log plot in
Fig. \ref{fig:peakheight_L}. From calculations for different chain lengths
$L=50$, $100$, $150$, $200$ we find values for $\alpha$ much closer to the
expected value.

\begin{figure}
\begin{center}
\epsfig{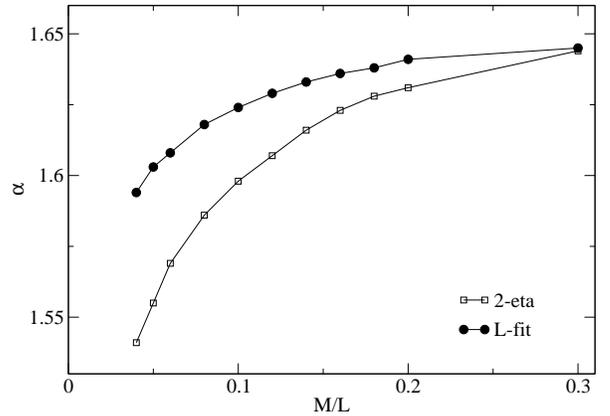}
\end{center}
\caption{Edge exponent $\alpha$ obtained from fitting the height of the first
peaks in the spectral function versus the chain length by a power law (full
circles). The open squares were obtained using
Eq.\ref{dsf-low}.}\label{fig:alpha}
\end{figure}

The values obtained from this $L$-fit and the expectations from $\alpha =
2-\eta$ are compared in Fig.~\ref{fig:alpha}. The agreement of both values is
good for all $M/L$ (deviation $<5\%$) and gets better for larger magnetizations.
Having found a reliable way to extract $\alpha$ we turn to the high-energy
continuum in $S^{zz}(k=\pi, \omega)$.

\begin{figure}
\begin{center}
\epsfig{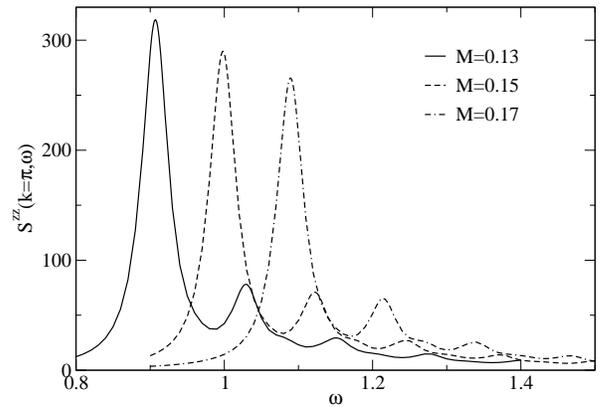}
\end{center}
\caption{$S^{zz}(k=\pi, \omega)$ for different magnetizations. Here
  the numerical broadening is $\delta=0.02$ and $m=500$ 
states were kept in the reduced DMRG basis.}\label{fig:upper_cont}
\end{figure}

Typical spectra for different magnetizations are shown in
Fig.~\ref{fig:upper_cont}.  The edge frequency shifts to higher energies with
increasing magnetization. Due to the large energy the convergence of the
algorithm is worse than for the previous continuum resulting in a significantly
larger DMRG basis. Apart from the expected series of peaks these spectra show
small additional peaks between each two peaks. These are finite-size artifacts
that vanish for sufficiently large system sizes.

\begin{figure}
\begin{center}
\epsfig{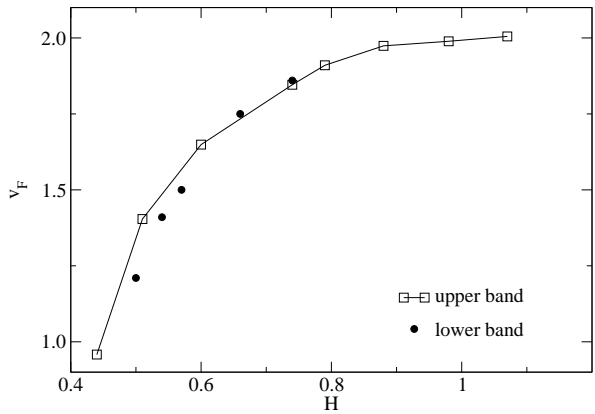}
\end{center}
\caption{Fermi velocities extracted from the low-energy spectrum (filled
  circles) and from the
high energy one (open squares).}\label{fig:v_Fermi}
\end{figure}

The Fermi velocity extracted from the upper continuum spectra is very similar to
the one obtained from the low-energy continuum as shown in
Fig. \ref{fig:v_Fermi}.

\begin{figure}
\begin{center}
\epsfig{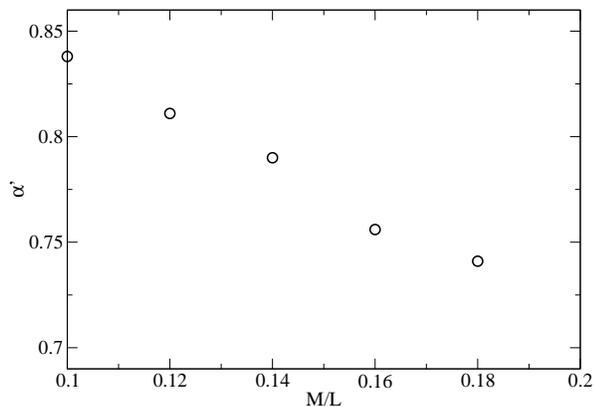}
\end{center}
\caption{The edge exponent of the high-energy continuum $\alpha'$ obtained from
the $L$-fit of the first peak height for different
magnetizations.}\label{fig:alphaprime}
\end{figure}

Like in the 
low-energy continuum, the peaks get broader with increasing energy,
making direct power-law fit of peak intensity vs frequency difficult. 
Thus, for determining the
edge exponent $\alpha'$ the same procedure based on the size
dependence is used 
as in the previous case. From those fits we find the exponent as shown in 
Fig.~\ref{fig:alphaprime}.

\begin{figure}
\begin{center}
\epsfig{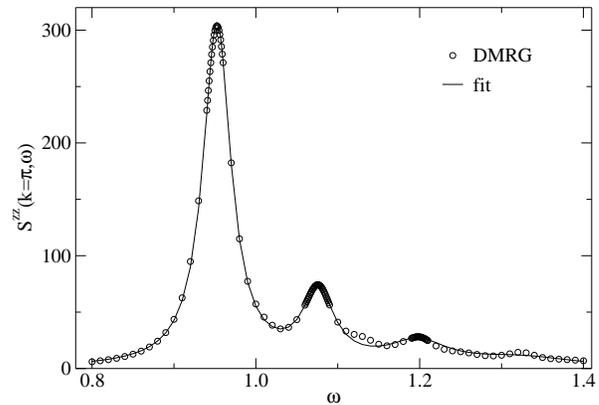}
\end{center}
\caption{Circles: $S^{zz}(k=\pi, \omega)$ obtained with $m=500$ states in the
DMRG basis.  Solid line: Fit by a sum of Lorentzians. For fitting parameters see
text. }\label{fig:upper_cont_fit}
\end{figure}

As a furher check for the consistency of our data, we show a fit of the high
energy continuum for a magnetization per site $M/L=0.14$ calculated in a $L=100$
site chain. We have fitted the spectrum to a sum of Lorentzians with different
widths and with a power-law decay of the peak intensity with the exponent fixed
at the value $\alpha'=0.79$ obtained from the analysis of the size dependence of
the peak heights.  The peak widths and the true edge frequency $\omega_0$ were
used as fitting parameters. The result is shown in
Fig.~\ref{fig:upper_cont_fit}. Since the small shoulder at $\omega \approx 1.13$
is a finite size effect it has not been fitted.  The values obtained for the
peak widths $\gamma_i$ are $\gamma_1 = 0.0207$, $\gamma_2 = 0.0236$, $\gamma_3 =
0.0442$, and $\gamma_4 = 0.0820$. These widths and the independently obtained
$\alpha'$ provide an excellent fit. The true edge frequency $\omega_0 = 0.929$
is slightly smaller than the position of the first peak $\omega_1 = 0.952$. To
check the validity of this particular value we perform an $L\rightarrow\infty$
extrapolation of $\omega_1(L)$ which is supposed to yield the edge frequency. We
find $\omega_1(L\rightarrow\infty) = 0.935$ which is slightly larger than the
one found from the fitting procedure.

\begin{figure}
\begin{center}
\epsfig{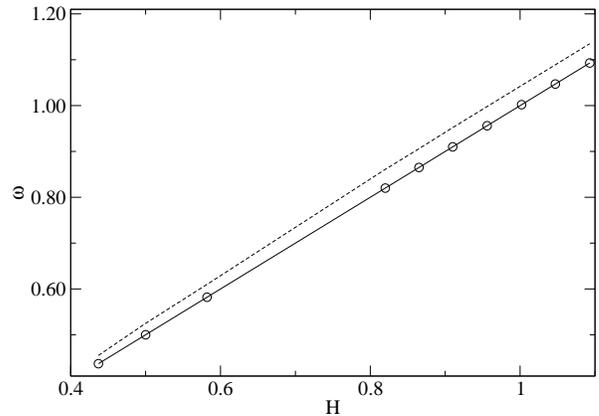}
\end{center}
\caption{Position of the first peak as function of the magnetic field (circles). 
The dashed line denotes $\omega_0=H$ as expected for the idealized
hardcore boson model.\cite{KM02prb,KM02ptp}}\label{fig:omega_H}
\end{figure}

Notably, the exponent $\alpha'$ decreases with increasing magnetization unlike
the edge exponent of the low-energy continuum.  With the knowledge of the edge
exponent $\alpha'$, the Fermi velocity $v_{F}$ and the static exponent $\eta$ we
can extract the effective ``'host-impurity'' interaction $U_f$ using Eqs.\
(\ref{eta1}), (\ref{dsf2}). The  values of $U_f$ extracted in that way lie 
between $7.7$ and $8.3$, which can be deemed approximately independent of the
magnetization (within $\pm 5\%$ accuracy).

The position of the first peak $\omega_{1}$ (which serves as a crude
approximation for the true edge frequency $\omega_{0}$) as a function of the
magnetic field shows a clear linear dependence (see Fig.~\ref{fig:omega_H}). It
should be noted that a linear dependence $\omega_{0}=H$ has been obtained in the
idealized hardcore bosonic model,\cite{KM02prb,KM02ptp} as well as in a
bosonisation calculation\cite{FurusakiZhang99} for a $S=\frac{1}{2}$ ladder.
The observed dependence is quite close to $\omega_{0}=H$ as seen in Fig.\
\ref{fig:omega_H}.

\section{Summary}

We have shown that the high-energy spectrum of a $S=1$ Heisenberg chain in a strong
 external magnetic field $H$ exceeding the Haldane gap $\Delta$ contains edge
 singularities, similar to those known to exist in the low-energy spectral
 response. It is found that in the frequency range $\omega\gtrsim \Delta$ the
 longitudinal (with respect to the applied field) dynamical structure factor is
 dominated by the power-law singularity $S^{\parallel}(q=\pi,\omega)\propto
 (\omega-\omega_{0})^{-\alpha'}$.  It is shown that the edge exponent of the
 high-energy continuum $\alpha'$ decreases with the magnetic field, consistent
 with theoretical expectations.  The edge frequency $\omega_{0}$ is found to
 increase linearly with the magnetic field, $\omega_{0}\simeq cH$, the
 coefficient $c$ being close to $1$, which agrees well with the predictions of
 Refs.\ \onlinecite{KM02prb,KM02ptp,FurusakiZhang99}. 
 The existence of power-law continua with edge singularities at high
 energies should be a quite general feature, common for all one-dimensional
 gapped spin systems in high magnetic fields and directly related to the
 Tomonaga-Luttinger liquid character of the ground state at $H>\Delta$.

\acknowledgments

We thank H.-J.~Mikeska for fruitful discussions which initiated the
present study. AF and US acknowledge support by the Deutsche
Forschungsgemeinschaft (DFG) under project DFG-Scho 621/4-1.
AK is supported by the grant KO~2335/1-1 under the Heisenberg Program
of the DFG.


\begin{thebibliography}{99.}

\bibitem{Tomonaga-Luttinger} S. Tomonaga, Prog. Theor. Phys. \textbf{5}, 544
  (1950);
J. M. Luttinger, J. Math. Phys.\textbf{4}, 1154 (1963).

\bibitem{SachdevSenthilShankar94} S. Sachdev, T. Senthil, and
R. Shankar, Phys. Rev. B \textbf{50}, 258 (1994).

\bibitem{Schulz86} H. J. Schulz, Phys. Rev. B \textbf{34}, 6372 (1986).

\bibitem{TL-nanotubes1} M. Bockrath, D. H. Cobden, J. Lu,
  A. G. Rinzler, R. E. Smalley, L. Balents, P. L. McEuen, Nature
  \textbf{397}, 598 (1999);
R. Egger, A. Bachtold, M.S. Fuhrer, M. Bockrath, D.H. Cobden,
P.L. McEuen, Lecture Notes in Physics \textbf{579}, 125 (2001).

\bibitem{TL-nanotubes2} H. Ishii, H. Kataura, H. Shiozawa, H. Yoshioka,
H. Otsubo, Y. Takayama, T. Miyahara, S. Suzuki, Y. Achiba, M. Nakatake,
T. Narimura, M. Higashiguchi, K. Shimada, H. Namatame, and M. Taniguchi, Nature
\textbf{426}, 540 (2003).

\bibitem{Wen90} X. G. Wen, Phys. Rev. B \textbf{41}, 12838 (1990).

\bibitem{Stone90} M. Stone, Phys. Rev. B \textbf{42}, 8399 (1990).

\bibitem{Haldane80-81} F. D. M. Haldane, Phys. Rev. \textbf{45}, 1358 (1980). 

\bibitem{Haldane83} F. D. M. Haldane, Phys. Rev. Lett. \textbf{50},
  1153 (1983).

\bibitem{AffleckHaldane87} I. Affleck, F. D.~M. Haldane,
  Phys. Rev. B. \textbf{36}, 5291 (1987).

\bibitem{ChitraGiamarchi97} R. Chitra and T. Giamarchi, Phys. Rev. B
  \textbf{55}, 5816 (1997).

\bibitem{KonikFendley02} R. M. Konik and P. Fendley, Phys. Rev. B
  \textbf{66}, 144416 (2002).

\bibitem{Campos02} L. Campos Venuti, E. Ercolessi, G. Morandi, P. Pieri and M. Roncaglia,
Int. J. Mod. Phys. B \textbf{16}, 1363 (2002).

\bibitem{Fath2003} G. Fath,  Phys. Rev. B  \textbf{68}, 134445 (2003).

\bibitem{Regnault+06} L.-P. Regnault, A. Zheludev, M. Hagiwara,
  A. Stunault, e-print cond-mat/0602357.

\bibitem{Hagiwara+05} M. Hagiwara, L. P. Regnault, A. Zheludev,
 A. Stunault, N. Metoki, T. Suzuki, S. Suga,
 K. Kakurai, Y. Koike, P. Vorderwisch, and J.-H. Chung,
 Phys. Rev. Lett. \textbf{94}, 177202 (2005).

\bibitem{Hagiwara+05a} M. Hagiwara, H. Tsujii, C. R. Rotundu,
  B. Andraka, 
Y. Takano, N. Tateiwa, T. C. Kobayashi, T. Suzuki, and S. Suga,
  e-print cond-mat/0511004.

\bibitem{Izumi+03} K. Izumi, T. Goto, Y. Hosokoshi, J.-P. Boucher. Physica
B \textbf{329-333}, 1191 (2003).

\bibitem{Yoshida+05} Y. Yoshida, N. Tateiwa, M. Mito, T. Kawae, K. Takeda,
Y. Hosokoshi, and K. Inoue, Phys. Rev. Lett. \textbf{94}, 037203
(2005).

\bibitem{Orendac+99} M. Orend{\'a}\v{c}, S. Zvyagin, 
A. Orend{\'a}{\v{c}}ov{\'a}, M. Seiling, B. L{\"u}thi, A.~Feher,
M.~W.~Meisel, Phys. Rev. B \textbf{60}, 4170 (1999)

\bibitem{Hagiwara+03} M. Hagiwara, Z. Honda, K. Katsumata,
A. K. Kolezhuk, and H.-J. Mikeska, Phys. Rev. Lett. {\textbf
91},  177601 (2003).

\bibitem{KM02ptp} A. K. Kolezhuk, H.-J. Mikeska, 
 Prog. Theor. Phys. Suppl. \textbf{145}, 85 (2002).

\bibitem{KM02prb} A. K. Kolezhuk, H.-J. Mikeska: Phys. Rev. B
\textbf{65}, 014413 (2002).

\bibitem{FurusakiZhang99} A. Furusaki and S.-C. Zhang: Phys. Rev. B \textbf{60},
  1175 (1999)

\bibitem{SorellaParola96} S. Sorella and A. Parola,
  Phys. Rev. Lett. \textbf{76}, 4604 (1996).

\bibitem{NetoFisher96} A. H. Castro Neto and M. P. A. Fisher, Phys. Rev. B
  \textbf{53}, 9713 (1996).

\bibitem{Tsukamoto98} Y. Tsukamoto, T. Fujii, and N. Kawakami,
  Phys. Rev. B \textbf{58}, 3633 (1998).

\bibitem{Sato06} M. Sato, J. Stat. Mech., P09001 (2006).

\bibitem{Affleck05} I. Affleck, arXiv:cond-mat/0508354.

\bibitem{SachdevTroyerVojta01} S. Sachdev, M. Troyer, and M. Vojta,
  Phys. Rev. Lett. \textbf{86}, 2617 (2001).

\bibitem{remark-on-current} Here we are interested only in the dynamical
  structure factor at the wave vector $q=\pi$ which corresponds to the
  ``impurity'' with zero momentum. In case of nonzero momentum an additional
  term of the type $iV_{f}^{a}\int dx\, \big( b^{\dag}\partial_{x}
  b-(\partial_{x}b^{\dag})b \big) (\partial_{x}\theta)$, which is a product of
  currents corresponding to $a$- and $b$-particles, could be present in the
  forward scattering.\protect{\cite{Tsukamoto98}} This term can be dealt with in
  a similar way,\protect{\cite{Tsukamoto98,Fiete06}} leading to a wave vector
  dependent edge exponent.

\bibitem{AffleckLudwig94} I. Affleck and A. W. W. Ludwig, J. Phys. A:
  Math. Gen. \textbf{27}, 5375 (1994).

\bibitem{White1992-11} S. R. White, Phys. Rev. Lett. \textbf{69}, 2863 (1992).

\bibitem{Schollwoeck2005} U. Schollw\"{o}ck, Rev. Mod. Phys. \textbf{77}, 259 (2005).

\bibitem{Kuehner1999} T. D. K\"{u}hner and S. R. White, Phys. Rev. B
\textbf{60}, 335 (1999).

\bibitem{VerstraeteMPS} F. Verstraete, A. Weichselbaum, U. Schollw\"ock,
J. I. Cirac, and Jan von Delft, arXiv:cond-mat/0504305 (unpublished).

\bibitem{White92} S. R. White, Phys. Rev. Lett. \textbf{69}, 2863 (1992).

\bibitem{Fiete06} G. A. Fiete, Phys. Rev. Lett. \textbf{97}, 256403 (2006).



\end{thebibliography}
\end{document}